\documentclass[12pt]{article}

\usepackage{amssymb}

\begin{document}

\begin{titlepage}

\begin{flushright}
\end{flushright}
\vskip 2.5cm

\begin{center}
{\Large \bf Astrophysical Bounds on the Photon\\
Charge and Magnetic Moment}
\end{center}

\vspace{1ex}

\begin{center}
{\large Brett Altschul\footnote{{\tt baltschu@physics.sc.edu}}}

\vspace{5mm}
{\sl Department of Physics and Astronomy} \\
{\sl University of South Carolina} \\
{\sl Columbia, SC 29208 USA} \\

\end{center}

\vspace{2.5ex}

\medskip

\centerline {\bf Abstract}

\bigskip

If the photon possessed an electric charge or a magnetic moment, light waves
propagating through magnetic fields would acquire new quantum mechanical phases.
For a charged photon, this is an Aharonov-Bohm phase, and the fact that we can
resolve distant galaxies using radio interferometry indicates that this phase must
be small. This in turn constrains the photon charge to be smaller that
$10^{-32}$ $e$ if all photons have the same charge and smaller than $10^{-46}$ $e$
if there are both positively and negatively charged photons.
The best bound on the magnetic moment comes from the observed absence of
wavelength-independent photon birefringence. Birefringence measurements, which
compare the relative phases of right- and left-circularly polarized waves, restrict
the magnetic moment to be less than $10^{-24}$ $e$ cm. This is just a few orders of
magnetude weaker than the experimental bounds on the electron and neutron electric
dipole moments.

\bigskip

\end{titlepage}

\newpage

\section{Introduction}

So far as we know, photons---quanta of the electromagnetic field---have zero
mass, zero charge, and zero magnetic moment. These three properties are all crucial
to our understanding of electromagnetism; they imply that the purely electromagnetic
sector of the standard model is scale invariant and noninteracting.

However, it is still interesting to ask how well we really know that all these
quantum numbers are zero. This can be a tricky question to ask
just about the photon mass,
but it is quite a bit trickier for the photon charge and magnetic moment, which
control the coupling of the electromagnetic field to itself. There are at least three
fully dynamical theories that can explain a photon mass---the Proca~\cite{ref-proca},
Higgs~\cite{ref-englert,ref-higgs,ref-guralnik}, and
Stueckelberg~\cite{ref-stueckelberg} models---but no reasonable theory of a
self-interacting photon has
ever been proposed. There are, of course, theories with interacting gauge
bosons---the non-Abelian gauge theories. However, the existence of a non-Abelian
gauge symmetry immediately implies the existence of a multiplet of vector bosons with
different charges. The charges in these theories are also quantized. The fact that
there is only one known type of photon, whose charge must be many orders of
magnitude smaller than the standard quantum of charge (the proton charge $e$), rules
out non-Abelian gauge theories as viable theories of a self-interacting photon.

There are more immediate physical differences between a photon mass and charge as
well. The mass parameter
in the Lagrangian of a Proca, Higgs, or Stueckelberg theory has meaning at the
classical level; it describes the the gap in the dispersion relation for classical
electromagnetic waves. A photon charge or magnetic moment, on the other hard, is
intrinsically quantum mechanical in nature. The electromagnetic couplings of a
propagating wave are crucially tied to the decomposition of that wave as a collection
of quantized photons.

Because a
putative photon mass is easier to accommodate than a photon charge or dipole
moment, much more attention has been paid to it. It is also a great
deal easier to place bounds on a quantity (like the mass) for which a complete
description of the physics exists.
In 2006, the Particle Data Group~\cite{ref-PDG} listed only four bounds on the photon
charge, compared with fifteen on the photon mass. There were no bounds listed for the
photon's dipole moment, even though the photon has an intrinsic spin. Yet the
photon charge and magnetic moment are at least sufficiently interesting that we would
like to know how tightly they can really be constrained.

However, the absence of a complete theory describing electrically or magnetically
interacting photons complicates the task of placing bounds on the strength of such
putative interactions. Again, this is in contrast with the case of a photon mass,
which can be bounded in many different ways.
The dynamical stability of magnetized galaxies gives the strongest
limit on the photon mass parameter~\cite{ref-chibisov}, although there are many
assumptions associated with
such a bound. The simple existence of a galaxy-scale magnetic field places limits
on a possible Proca mass for the photon, and a large ambient vector
potential associated with such a field could give rise to laboratory-measureable
torques~\cite{ref-lakes,ref-luo}.
More secure bounds, which do not require any inferences about fields on galactic
scales, come from observations of very low frequency
magnetohydrodynamic waves in nebulae~\cite{ref-barnes}, direct tests
looking for the screening of Earth's~\cite{ref-fischbach} or
Jupiter's~\cite{ref-davis} magnetic field, and
measurements combining screening of the solar field and magnetohydrodynamic
effects~\cite{ref-ryutov}.

These tests are possible because we can interpret
the results of our observations in the context of a well-defined theory.
This is usually taken to be the Proca theory; there are critical differences if the
photon mass arises from spontaneous symmetry breaking~\cite{ref-adelberger}.
Without a complete theory describing the photon charge, we must rely on rather
different techniques. For example, while positing the existence of a photon charge
does tell us things about how photon propagation must be affected, it does not tell
us anything reliable about how static electromagnetic fields will be modified.

In fact, the question of static fields in a charged photon theory is rather
problematic. A macroscopic field can be
``built up'' from photons; yet even in strong fields, there is no evidence that the
electromagnetic field carries charge. One can get around this by supposing that
photons with multiple charges exist---that there are both positively and negatively
charged photons and possibly neutral ones as well. Whether this is actually an
improvement in the situation, however, is a matter of opinion. Having photons with
different charges makes it possible to build up macroscopic fields that obey charge
neutrality, but it also means that photons with different charges must be able to
interfere, since experimentally, photons have never been observed not to interfere.
One cannot avoid this
by claiming that only one type of photon is ever actually being observed, the others being somehow invisible and not interacting with our detectors, because what we mean
by a photon is really just the electromagnetic quantum that we can observe. If there
are other quanta with different charges that we do not see, they are not photons in
any meaningful sense.

Having interference between particles of unequal charges violates local charge
conservation and charge superselection. However, this may not be such a serious
difficulty. Any new theory that contains charged photons is unlikely to preserve
gauge invariance. The Proca and Higgs theories describing photon masses involve
either explicit or
spontaneous breaking of gauge symmetry. It is easy to believe that a charged photon
theory will lose gauge invariance as well.

However, having both positive and negative photons is potentially
problematical for another reason as well.
Assuming that these charged particles are massless, photon pairs could be produced at
no energy cost. This could lead to complete screening of charges. Such screening
does occur in theories with massless charged fermions~\cite{ref-gribov}; charges are
confined in neutral meson-like excitations. It seems reasonable that this might be a
general feature of theories featuring massless charges.

For a long time,
the best laboratory bound on the photon charge came from an old experiment looking
for evidence of fractional charges~\cite{ref-stover}. The experimental setup had a
light beam shining on an iron
spheroid and an arrangement for monitoring the charge on the spheroid. The photon
charge bound (which was largely incidental)
was derived by dividing the total change in the spheroid's charge over
an extended experimental run by the number of photons absorbed by the spheroid
during that time. This seems a straightforward enough procedure, but it has a serious
deficiency. According to quantum electrodynamics,
during any photon absorption event, a charged particle will emit a large
number of extremely low frequency bremsstrahlung photons. Formally, the number
emitted is infinite, limited only by an infrared regulator~\cite{ref-bloch}. This
fact is, in itself,
another problem for theories with charged photons, since our current understanding
indicates that every photon interaction event increases the number of propagating
photons present by an infinite number. To get around the infinity, we would expect
that the bremsstrahlung should be drastically modified at very low frequencies by
the presence of even a very small photon charge; at sufficiently low frequencies,
the electromagnetic interaction energy is going to dominate the usual photon energy.
But at higher yet still quite small frequencies, the bremsstrahlung emission is a
real effect, which the experimental bounds do not take into account. This casts this
particular laboratory result into serious doubt, and it illustrates the difficulty
associated with trying to measure a quantity without a full theory describing it.

Careful consideration of the preceding laboratory experiment is also useful for
another reason. The bound on the photon charge derived from that experiment, which
was at the $10^{-16}$ $e$ level, gives an indication of what ranges of photons
charges could be
considered ``large'' and ``small.'' The $10^{-16}$ $e$ bound indicates what
magntiude the photon charge would have to have if the total number of photons in the
experiment were to have an
aggregate charge that would be observable by macroscopic means.
This is a very rough dividing line between the ``large'' and ``small'' charge
regimes. Because photons are so numerous, they can have individual charges that are
many orders of magnitude smaller than $e$, yet still carry a signficant charge in
bulk.

Any bound on the photon charge or magnetic moment is going to be somewhat
uncertain, but some bounds are more robust that others. It is quite easy to show that
the photon charge is very, very small. One can simply measure the change in a
photon's energy between a source and detector placed at different voltages. This is
a suitable experiment for an undergraduate laboratory course, and using M\"{o}ssbauer
spectroscopy, students may place bounds at the $\sim10^{-10}$ $e$ level in only a few
hours.

More sophisticated bounds on the photon charge have often been based on the
observation that a charged photon would be deflected by a magnetic field. The photons
and fields involved may be either astrophysical~\cite{ref-cocconi1,ref-kobychev}
or created in the laboratory~\cite{ref-semertzidis},
and the best bound arrived at using this
strategy is at the $4\times 10^{-31}$ $e$ level~\cite{ref-kobychev} if all photons
have the same charge and the $3\times 10^{-33}$ $e$ level if multiple charges exist.
Also associated with the deflection of charged photons is a frequency-dependent time
delay for photons reaching Earth; less energetic photons from a given source will be
deflected onto longer arcing trajectories. The best bound inferred from the sharpness
of pulsar pulses is $5\times 10^{-30}$ $e$, reported in~\cite{ref-raffelt} (which
corrects an error
in~\cite{ref-cocconi2}). All the astrophysical bounds require information about
magnetic fields in space.

If we observe photons coming from a single source over a range of energies
$\Delta E$ centered around $E$, and the angular spread of these photons is $\Delta
\theta$ after
they have traveled a distance $L$ through a magnetic field $B$ perpendicular to their
path, we can constrain the photon charge $q$ to be
\begin{equation}
\label{eq-deflect}
\frac{|q|}{e}<\frac{2E^{2}\Delta\theta}{eBL\Delta E}.
\end{equation}
If extragalactic field configurations are used, $BL$ would be replaced by
something of order $B\sqrt{L\lambda_{C}}$, where $\lambda_{C}$ is the correlation
length of the field.
When this bound on $q$ is expressed in terms of the photon energy
(as opposed to frequency), the result is independent of $\hbar$.
The deflection decreases with the photon energy, since higher-energy
photons possess more momentum and are thus less deflected by the energy-independent
Lorentz force. The bound (\ref{eq-deflect}) applies to photons all of the same
charge. If there are both positively and negatively charged photons, they will be
deflected in opposite directions, leading to a more sensitive bound that does not
depend on the photons having a finite spread in energy. Instead, the bound is
\begin{equation}
\label{eq-deflect2}
\frac{|q|}{e}<\frac{E\Delta\theta}{eBL}.
\end{equation}

Indirect searches for the effects of a photon charge are also possible. Measurements
of the anisotropy of the cosmic microwave background (CMB) can be used to place
strong constraints on the physics of the early universe. In the radiation dominated
era, a photon charge could make huge contributions to the energy density, but no
signature of this is seen in the CMB~\cite{ref-sivaram}. This leads to bounds on the
photon charge, which
depend somewhat on how the conductivity of the early epoch is modeled. If there are
no cancellations between different species, and a nonzero photon charge disturbed the
overall charge neutrality of the early universe, the best limit on the photon charge
is at the $10^{-35}$ $e$ level~\cite{ref-caprini}. Importantly, there are no CMB
bounds for the case of multiple photon charges. In that case, the photon gas remains
neutral, and there is no buildup of potential energy when many photons are crammed
together.

The purpose of this paper is to explore new techniques for bounding the photon charge
and the previously little discussed photon magnetic moment. We can place bounds on
the charge using the quantum mechanical Aharonov-Bohm effect. Charged particles moving along different paths through a magnetic field
pick up different phases, and the observed coherence of photons from distant
astrophysical sources allows us to place bounds on this effect and hence on the
photon charge~\cite{ref-altschul8}.
Uncharged but magnetized photons traversing different paths though an inhomogeneous
magnetic field will likewise acquire phase differences, leading to analogous bounds.
The photon magnetic moment can also be bounded by looking at photon
birefringence. Photons that have
traversed cosmological distances can be very precise probes of
novel phenomena in electrodynamics. The immense distances over which they
travel magnify miniscule effects. After millions of parsecs, a tiny change in how electromagnetic waves propagate can give rise to a readily observable effect.

Obviously, the notions that the photon could have charge or that it could possess a
magnetic moment are related. Any particle with both charge and spin would naively
be expected to possess a nonzero magnetic moment as well. However, the typical
magnitude of the magnetic moment for a particle of charge $q$ and mass $m$ is
$q\hbar/mc$, which is infinite for a massless particle like the photon. So
while we expect that a charged photon could well also have a magnetic dipole moment,
there is no natural relationship between the size of the particle's electric and
magnetic couplings. Because of this, and because we expect both couplings to be
small, we shall treat them one at a time, although a combined treatment allowing for
both would be straightforward.
We shall discuss the interferometric bounds on the photon charge in
Section~\ref{sec-charge} and on the magnetic moment in Section~\ref{sec-muint}.
In Section~\ref{sec-biref}, we shall
show that we can place significantly better bounds on the dipole moment by
measuring phase differences in another way---via photon birefringence.
Section~\ref{sec-concl} offers some concluding discussion.

\section{Interferometric Bounds: Charge}
\label{sec-charge}

We would like to place bounds that do not
depend in any crucial way on the intricate details of the interacting
photon dynamics. We shall assume only that there exists an effective Lagrangian
$L_{{\rm eff}}$
governing the propagation of a single photon. The coupling of a charge $q$ photon
to an external electromagnetic field should take the form $L_{q}=
-\frac{q}{c}v_{\mu}A^{\mu}_{{\rm ext}}$, where $v^{\mu}=(c,c\hat{v})$ is the photon's
four-velocity. This Lagrangian is essentially unique, once we specify that there must
be a potential energy term $-qA^{0}_{{\rm ext}}$ and demand conventional Lorentz
transformation
properties. The equation of motion derived from $L_{q}$ is the Lorentz force law.

Associated with $L_{q}$ is an additional phase that a charged photon
picks up as it travels, relative to a conventional uncharged photon.
If we take the eikonal
approximation, in which the photon's deflection from a straight-line path is
neglected, this phase is
\begin{equation}
\phi=\frac{1}{\hbar}\int_{0}^{t}d\tau\, {\cal L}_{q}.
\end{equation}
The time interval of the photon's flight is from 0 to $t$. Assuming the main
contribution to be magnetic, we neglect the effects
of the electrostatic potential. Then, taking the total
distance traveled to be $L$, the phase is
\begin{equation}
\label{eq-phase}
\phi=\frac{q}{\hbar c}\int_{0}^{L}d\vec{\ell}\cdot\vec{A}_{{\rm ext}}.
\end{equation}

The phase $\phi$ depends on the path, and we would like to compare it for two rays
emanating from the same source.
We can do this by observing the phase difference between photons arriving at
different points. This is the basis of
astrophysical interferometry. If two telescopes, separated by a
baseline $d$, collect data from a source lying approximately in the plane
perpendicular to the baseline, the observed phase difference due to a possible photon
charge is equal to the difference between two phases $\phi_{1}$ and $\phi_{2}$ of the
form (\ref{eq-phase}). Neglecting a miniscule contribution
proportional to the integral of $\vec{A}_{{\rm ext}}$ along the baseline, the
phase difference is $\Delta\phi=\Phi q/\hbar c$, where $\Phi$ is the magnetic flux
threading between the two lines of sight. This is the standard Aharonov-Bohm phase
difference, which is independent of the photon energy.

To estimate the flux $\Phi$, we must know something about the relevant magnetic
fields. For randomly oriented fields, with typical magnitude $B$ and correlation
length $\lambda_{C}$, $\Phi$ is proportional to $Bd\sqrt{\lambda_{C}L}$. We would
like to get an idea of the accompanying numerical constant. To do this,
we assume that the line of sight
passes though $L/\lambda_{C}$ magnetic field domains, each of equal size. In each
domain, the field is randomly oriented along one of the six cardinal directions;
therefore
only one third of the domains contribute to the total flux. The average value of
$\Phi$ is then determined by the statistics of a one-dimensional random walk with
$L/3\lambda_{C}$
steps. The mean distance from the origin after $L/3\lambda_{C}$ steps
is $\sqrt{2L/3\pi\lambda_{C}}$. To get $\Phi$, we multiply this by a flux
$Bd\lambda_{C}$ and a factor of $\frac{1}{2}$
corresponding to the triangular geometry of the threaded region. The total flux
is then
\begin{equation}
\Phi\approx\sqrt{\frac{L\lambda_{C}}{6\pi}}dB.
\end{equation}
The precise numerical constants are not all that important. The main point is to
show that they are not such as to change drastically the order of magnitude of the
result.
Including cosmological effects and the expansion of the universe tends to change the
flux relatively little, the shortening of the true path length being compensated for
by the stronger fields [$B$ proportional to a positive power of $(1+z)$] in the early
universe~\cite{ref-ryu}.

If photons moving along different paths acquired Aharonov-Bohm phase differences,
this would eventually interfere with interferometry. In order for
interferometry to be possible, photons arriving at different telescopes must have
definite phase relations. A $\Delta\phi$ of order 1 or larger would destroy these
necessary relations. The decoherence that would be caused by a nonzero $\Delta\phi$
has never been seen, and this leads directly to a limit on the photon charge.

There is a possible objection that the Aharonov-Bohm phase, being essentially quantum
mechanical, should not contribute to the
ordinary, essentially classical, phase that we can observe with radio waves.
In this view,
the novel phase would represent some kind of intrinsically quantum effect, perhaps
only observable if a single photon were bifurcated and recombined (whereas in radio
interferometry, we observe distinct but coherent photons
at different locations). This would have to represent a new kind of interference, entirely separate from the usual interference of electromagnetic waves.
Without a viable theory of self-interacting photons, we cannot reject such a proposal
automatically; however, it seems even more farfetched than the possibilities of
photon charges and magnetic moments. The proposal violates the usual relationship
between photons and classical waves; according to the correspondence principle, the
classical and photon phases are identical.

Of course,
there are other phase uncertainties in real interferometry. Telescopes'
relative positions are known to limited accuracy, but position uncertainties lead to
phase differences that are proportional to the photon frequency, and they have a
predictable dependence on the direction of observation. This is in contrast with
an Aharonov-Bohm phase difference, which is frequency independent and varies randomly
with different pointing directions. To overcome the real phase uncertainties,
sophisticated fringe finding algorithms are required, and individual telescopes are
often calibrated by observing reference sources located close to the real sources of
interest. The reference sources tend to be comparatively nearby, however, so photons
arriving from them will not have the same Aharonov-Bohm phases as photons coming from
the same direction but emanating from distant galaxies. Moreover, the
Very Long Baseline Interferometry Space Observatory Program (VSOP)
experiment, which has the longest baselines of any available interferometer,
owing to its use of the Highly Advanced Laboratory for Communications and Astronomy
(HALCA) satellite, had to do much of its calibration by dead reckoning, because
adjusting the alignment of the satellite was too time consuming~\cite{ref-scott}.

Our ability to study objects
at a distance $L$ with interferometers of baseline $d$ limits the photon charge to
be smaller than
\begin{equation}
\label{eq-bound}
\frac{|q|}{e}<\sqrt{\frac{6\pi}{L\lambda_{C}}}\frac{\hbar c}{deB}.
\end{equation}
This bound, based as it is on the Aharonov-Bohm phase, cannot be expressed in an
$\hbar$-independent form.
Since the Aharonov-Bohm phase is independent of energy, this bound is also independent of the photon energy $E$, in contrast to (\ref{eq-deflect}).
However, it is still most advantageous to work with low-energy photons, because their
phases can be determined most accurately.

The most problematical quantities in the expression (\ref{eq-bound}) are those
that characterize the cosmic magnetic field---$B$ and $\lambda_{C}$. These are
tricky to estimate separately, but fortunately, (\ref{eq-bound}) depends on the
specific combination $B\sqrt{\lambda_{C}}$. Many potentially observable effects
depend on this particular combination of parameters and for essentially the same
reasons as in our bound. Weak magnetic effects tend to depend on the absolute value
of the time integrated field that a traveling particle interacts with, and we
already saw that this was proportional to $B\sqrt{L\lambda_{C}}$ for a trajectory of
length $L$.

The best upper bounds on extragalactic fields come
from observations of the Faraday rotation of photons moving through
plasmas~\cite{ref-vallee,ref-kronberg,ref-blase}. The precise bounds
one may derive from the Faraday observations depend on the assumptions one makes
about the large scale structure of the field and obviously on $\lambda_{C}$.
A bound of $B\lesssim10^{-8}$ G is reasonable, while cosmic ray and
high-energy photon data from the source Centaurus A suggest that $10^{-8}$ G may
also be an approximate lower bound for the magnetic field strength in the relative
vicinity of our galaxy~\cite{ref-anchordoqui}.
We shall use a rather low estimate of the extragalactic
magnetic field. Cosmic ray data suggest that $B\sqrt{\lambda_{C}}$ may be roughly at
the $10^{-10}$ G Mpc$^{1/2}$ level~\cite{ref-lemoine}. This estimate depends on
another conservative estimate of the number of ultra-high-energy cosmic
ray sources. The density of sources could quite reasonably be higher, leading to a
higher value of the field.

While $B\sqrt{\lambda_{C}}$ is an externally prescribed (if not entirely well known)
quantity, the parameters $d$ and $L$ are experimental
variables. The interferometry experiment with the largest values of $d$ is VSOP,
which makes use of the HALCA telescope in space as well as Earth-based observatories.
The resulting baselines extend up to $d>3\times 10^9$ cm. Using this large
interferometer, the VSOP experiment has studied active galactic nuclei out to
redshifts of $z\approx 3.5$.
To place a very conservative bound on $q$, we can choose the distance $L$ to be
1 Gpc, which is corresponds to a redshift less than 1. With this estimate of $L$,
along with $B\sqrt{\lambda_{C}}=10^{-10}$ G Mpc$^{1/2}$ and $d=3\times 10^9$ cm,
our conservative bound on the photon charge is
\begin{equation}
\label{eq-numeric}
\frac{|q|}{e}\lesssim 10^{-32},
\end{equation}
which is already the best direct bound for the case in which all photons have the
same charge. If we instead consider the most distant sources the VSOP experiment
studied (such as the quasar PKS 2215+020 at $z=3.57$) and make the same
assumption about cosmological fields as were made in~\cite{ref-kobychev}---that
$B\propto(1+z)^{2}$---the bound improves to
\begin{equation}
\label{eq-numeric2}
\frac{|q|}{e}\lesssim 1.5\times10^{-33}.
\end{equation}
A single demonstration that interferometry
is possible, using photons from just one source, places a bound at roughly the order
given. The fact that interferometry is possible for photons arriving from virtually
any direction indicates that these kinds of bound are quite robust.

It is also possible to use this technique to place bounds on the photon charge if
both $+q$ and $-q$ photons exist. In fact, the bounds are significantly better in
such a scenario.
Although the phase difference for two particles of equal charge is
gauge invariant, for particles with different charges it is not. Conventionally,
this is not a problem, since particles with different charges can never interfere.
However, if there are photons with different charges which do interfere, this
raises further questions.
Since the phase differences that we measure via such interference are not gauge
invariant, the gauge in this scenario must be fixed.
The gauge fixing condition ought to arise naturally out of the full theory describing
the photon charges. This would be analogous to the way that the
Lorenz gauge condition $\partial^{\mu}A_{\mu}=0$ must be satisfied if we introduce a
Proca mass term and insist on current conservation. Unfortunately however, we lack a
complete theory, and the precise form of the gauge condition is unknown.

Still, it is possible to place an order of magnitude bound on the charge,
assuming the large scale structure of the magnetic field is not modified. In a
magnetic field
domain of size $\lambda_{C}$, the typical vector potential is $B\lambda_{C}/2$.
Conservatively assuming that the vector
potential falls back to zero at the edge of the domain, there is a contribution
to the phase of a charge $q$ photon of  $\phi=\sqrt{\frac{L}{6\pi}}\frac{\lambda_{C}
^{3/2}qB}{\hbar c}$. The factor of $\sqrt{2L/3\pi\lambda_{C}}$ is the same as
before.
The phase difference for photons of charges $q$ and $-q$ is then
\begin{equation}
\Delta\phi=\sqrt{\frac{2L}{3\pi}}\frac{\lambda_{C}
^{3/2}|q|B}{\hbar c},
\end{equation}
which is independent of the baseline $d$. There is a phase difference even if two
photons follow exactly the same path, because they can have opposite charges and
hence pick up opposite phases.

The phase difference $\Delta\phi$ in the multiple charge case is not a systematic
phase difference between the phases observed at different points but rather a phase
uncertainty in the photons seen at a single observatory. If the observed photons have
equal probabilities of being positively or negatively charged (or positively,
negatively, or null charged), and the mean number of photons collected during a
given period is $\langle N\rangle$, the signal is subject to a phase uncertainty
proportional to $\Delta\phi/\sqrt{\langle N\rangle}$. This is in sharp contrast to
the essentially classical behavior that would ordinarily be seen in observations of
a first-order coherent photon beam. Radio interferometers routinely
make measurements in the regime where $\langle N\rangle$ for a reasonable observation
period is not too much larger than 1. Such measurements reveal no evidence of a
phase uncertainty that falls off only as $\langle N\rangle^{-1/2}$, seeing instead
a conventional $\langle N\rangle^{-1}$ uncertainty in the measured phase. This
indicates that $\Delta\phi$ is small.

In the resulting expression for a bound on $q$, the baseline in
(\ref{eq-bound}) is replaced with
a quantity proportional to $\lambda_{C}$, improving the constraint on the magnitude
of the charge by
a factor of ${\cal O}(d/\lambda_{C})$. $\lambda_{C}$ is more difficult to determine
than $B^{2}\lambda_{C}$, but choosing a relatively conservative value of 100 kpc
gives an improvement of ${\cal O}(10^{-14})$.
Taking our most conservative estimate of the distance to the source ($L\sim 1$ Gpc)
the bound on $q$ is
\begin{equation}
\label{eq-numstrong}
\frac{|q|}{e}\lesssim 10^{-46},
\end{equation}
while using sources farther away would again improve the constraint.
This is the best bound extant on a photon charge, and it applies to the multiply
charged case where the next best bounds (from the CMB) do not apply.

The VSOP experiment observed photons that had frequencies of 1.6, 5, and 22 GHz.
This places the energies of the photons from which our bounds on $q$ were derived in
the 6--90 $\mu$eV range at the time of their absorption. We might expect that the
photon charge should be independent of energy, as is the charge of other particles.
However, if the photon charge arises through the breaking of Lorentz symmetry,
something more exotic might be involved.

\section{Interferometric Bounds: Magnetic Moment}
\label{sec-muint}

The same kind of analysis can be applied to the possibility of the photon possessing
a magnetic moment.
In this case the phase involved is not an Aharonov-Bohm phase, but a slightly more
conventional dynamical phase. The interaction Lagrangian is $L_{\mu}=
\vec{\mu}\cdot\vec{B}_{{\rm ext}}$, where the magnetic moment is $\vec{\mu}=\mu
s\hat{k}$ for a photon of helicity $s$. $L_{\mu}$ does not have correct Lorentz
transformation properties on its own, but the extra terms needed to fix this problem
involve the electric field, which ought to have only negligible effects in deep
space.

The phase that a magnetized photon acquires as it travels is 
\begin{equation}
\phi=\frac{\mu s}{\hbar c}\int_{0}^{L}d\vec{\ell}\cdot\vec{B}_{{\rm ext}}.
\end{equation}
It is immediately obvious that this will give rise to a bound on $\mu$ very similar
to the bounds we have already derived for $q$. We need only replace $q$ with $\mu$
and $\vec{B}_{{\rm ext}}=\vec{\nabla}\times A_{{\rm ext}}$ with
$\vec{\nabla}\times B_{{\rm ext}}$ in the single charge result
(since the HALCA telescope only collected photons of one helicity). The
characteristic size of $\vec{\nabla}\times B_{{\rm ext}}$ is $B/\lambda_{C}$, so the
bound on $\mu$ that can be inferred from the fact that $\mu$ hasn't interfered with
astrophysical interferometry at a distance $L$ is
\begin{equation}
\label{eq-mubound}
|\mu|<\sqrt{\frac{6\pi\lambda_{C}}{L}}\frac{\hbar c}{dB}.
\end{equation}
If $\lambda_{C}\sim1$ Mpc (a conservatively large estimate) and $L\sim1$ Gpc, the
corresponding bound on $\mu$ is only at the $3\times 10^{-8}$ $e$ cm level. This is
not a very tight bound; it is more than a thousand
times larger than the Bohr magneton
$\mu_{B}$. A photon magnetic moment this large is presumably ruled out by atomic
experiments, where it could give rise to a large anomalous AC Zeeman effect. So
interferometry does not provide a particularly useful constraint on the magnetic
moment of the photon. However, there is a better way to place bounds on
$\mu$---using birefringence.

\section{Birefringence}
\label{sec-biref}

The relationship between the bound on $\mu$ derived from birefringence and the
bound derived from interferometry is analogous to the relationship between the
photon charge bounds in the multiple versus single charge scenarios. The
interferometric bounds on $\mu$ come from comparing the phases of waves originating
at the same source but following slightly different paths. Birefringence occurs when
there are two photon polarization states that interact differently
with the cosmic magnetic field, even while following the same path.
The resulting phase difference between right- and
left-circularly polarized photons can be measured directly, by looking at the change
in the polarization of linearly polarized waves, which are superpositions of the two
helicity states. This is analogous to using interferometry to compare the phases of
positively and negatively charged photons, which magnetic fields deflect in opposite
directions.

Birefringence in vacuum has been searched for and not seen. A systematic difference
between the phase speeds of positive and negative helicity photons does not
exist~\cite{ref-carroll1,ref-goldhaber,ref-wardle,ref-carroll2,ref-loredo}. The most
sensitive searches for this effect were done in the
context of a Lorentz-violating Chern-Simons modification of ordinary electrodynamics.
In that scenario, the phase speed difference between the two helicities was
independent of the magnetic field in the intervening space.
In contrast, if the photon possessed
a nonzero magnetic moment, we would expect to see phase differences that were random
(although with a well-determined characteristic size), since they would be determined
by the randomly oriented magnetic fields along the line of sight.
The phase difference would not depend on the
direction to the source in any predictable way, and this makes translation of the
known birefringence results slightly tricky.

The phase disparity between the left- and right-circularly polarized photons due to
a magnetic moment term is independent of frequency. This is an important property
that this kind of birefringence shares with Lorentz-violating Chern-Simons
birefringence. Photons moving through space definitely do experience birefringence,
but only because of the presence of free elections. The magnitude of the conventional
Faraday rotation is proportional to the photon wavelength squared, and consequently,
this effect can be subtracted away.

Searches for systematic differences in the phase speed between the two helicities
have looked at
the radiation from quasars with resolvable jets. The key quantity was the
angle between the jet direction and the plane of polarization the source's
synchrotron emission. If this angle depended on the distance to the source, that
would be strong evidence of birefringence, but no such dependence seems to
exist. Indeed, for high redshift sources, the
observed polarizations appear to be concentrated around the directions
normal to the jets in the plane of the sky, independent of
sources' distances. This kind of polarization is exactly what we would expect for
sources with magnetic flux lines pointing along their jets. Synchrotron electrons
revolve around these flux lines, emitting radiation that is polarized perpendicular
to the magnetic field and hence the jet. The fact that this angular
correlation persists even after photons have traversed cosmological distances
indicates that the photon's magnetic moment must be small. If $\mu$ were substantial,
then the two polarizations would acquire signficantly different phases over the
course of their propagation, and the radiation observed on Earth would be linearly
polarized in an effectively random plane.

The fact that the polarization is not randomized indicates that the
relative phase between
the left- and right-handed photons is less than 1. A more comprehensive analysis,
combining the information available from many sources, could presumably place
tighter bounds on the phase shift. This kind of analysis has been carried out as
part of the searches for Lorentz violation, but we shall not do it here. Instead,
we shall follow the same conservative procedures as we have previously used when
looking at interferometry data.

That the magnetic phase shift between two oppositely polarized photons traveling
along the same path is less than 1 is an indication that
\begin{equation}
\label{eq-mutight}
|\mu|<\sqrt{\frac{3\pi}{8L\lambda_{C}}}\frac{\hbar c}{B}.
\end{equation}
By comparing the phases of photons moving along the same path but interacting
oppositely with the magnetic field, we have again improved our bounds by a factor of
${\cal O}(d/\lambda_{C})$. Since we used a more conservative value of 1 Mpc in
evaluating (\ref{eq-mubound}), the improvement is now by a factor of
${\cal O}(10^{-15})$. The final result, again using a distance $L\sim1$ Gpc
(and the birefringence analyses have not extended out to redshifts as high as
those examined in VSOP interferometry), is
\begin{equation}
\label{eq-munumeric}
|\mu|\lesssim 10^{-24}\, e\, {\rm cm}.
\end{equation}

Obviously, this is a much stronger bound. Other dipole moments which are so far as
we know zero are constrained at comparatively similar levels. The neutron and
electron electric dipole moments (which violate CP)
are bounded at slightly better than the $10^{-25}$ $e$
cm~\cite{ref-harris,ref-altarev} and $10^{-27}$ $e$ cm~\cite{ref-regan} levels,
respectively.

Of course, it is from studies of photon birefringence that much of the best data on
astrophysical magnetic fields comes. The rotation measure (RM) of a source
characterizes
the change in polarization during light's transit due to the Faraday effect. So far,
measurements of RM have
not provided any direct evidence for extragalactic fields. This
might seem to indicate a problem with this technique for bounding $\mu$---using the
absence of birefringence in a magnetic field too weak to be measured by birefringence
to constrain an exotic effect. This certainly suggests that better knowledge of
cosmic magnetic fields is important for improving the bounds on $\mu$. We should
remember, however, that if birefringence due to the Faraday effect and/or a photon
magnetic moment were observed, the two would be straightforward to disentangle,
because of their different frequency dependences.

\section{Conclusion}
\label{sec-concl}

Radiation coming from distant galaxies often turns out to have features that would be
quite sensitive to exotic modifications of known physics.
When the hugeness of cosmic distance scales can be put to use, very tight
bounds of these modifications result. However, just how senstive
a photon measurement really will be may not be at all obvious until an actual
calculation is performed. 
The interferometric contraint on $\mu$ is actually quite poor; the bounding value
is several orders of magnitude larger than the Bohr magneton. Yet the
birefringence bounds are much better. The birefringence measurement looks
at the phase difference between the two helicities of photons, which
interact oppositely with the magnetic field. The interferometry bound comes from
comparing the phases of photons with the same helicity that have traversed different
paths and is worse by the large ratio $\lambda_{C}/d$, which represents
the fractional change in the extragalactic magnetic field over a distance equal to
the size of the interferometer. The large difference between the bounds on the
charge in the single and multiple charge scenarios arises in precisely the same
way. However, it was by no means obvious {\em a priori} that allowing for multiple
charges would lead to such a huge improvement in the bounds. For comparison, the
corresponding
improvement in the charge bounds due to light deflection is only about two orders of
magnitude; the improvement in that case is related to the replacement of the
potentially small parameter $\Delta E/E$ in (\ref{eq-deflect}) with the
constant 2.

The possibility of using Aharonov-Bohm phases to constrain the photon charge was
not noticed until quite recently. Since exotic and unlikely modifications of
known physics such as a photon charge or magnetic moment are little studied, it is
not really surprising that a promising method for placing bounds might be overlooked.
In fact, there may well be other comparatively straightforward ways to constrain the
self couplings of the electromagnetic field which have simply escaped researchers'
attention.

It is completely coincidental that, for the single charge case, the best bounds from
deflection and pulsar timing are comparable to those from the Aharonov-Bohm effect.
Improvements in both types of bounds are naturally possible. The deflection bounds
depend on the mean photon energy from a given source $E$ and the energy spread
$\Delta E$. Less energetic photons, measured over a wider range of
energies, will give stronger constraints. The measured angular spread $\Delta\theta$
is another quantity which might be improved experimentally.

The experimental variables in the interference bound are different.
The largest improvements in the single charge case might come from using longer
baselines. In principle, a baseline of 2 AU is available for certain types of
interferometric measurements, and a baseline this long would improve the bound on $q$
by four orders of magnitude.
However, doing interferometry using a single telescope and
measurements separated by half a year is obviously a daunting prospect
experimentally; it is not going to happen in the near future. A more reasonable
possibility for improvement in the short term---and one which is relevant in both the
single and multiple charge cases---involves correlating phase data from many
sources. The present bounds basically assume that only a phase decoherence $\Delta
\phi\sim 1$ for a small number of
distant sources is ruled out by the availability of
interferometric data. By combining data from multiple sources and baselines, it
should be possible to tighten the overall bound on $q$ somewhat, although it is
difficult to estimate quantitatively how much improvement is possible; this
procedure would also allow us to assign proper confidence levels to the bounds.

The quantities $L$, $\lambda_{C}$, and $B$ are determined by the sources we choose
to observe. A better understanding of magnetism on extragalactic scales will provide
more secure (but not necessarily numerically tighter) bounds on $q$. This applies
to the bounds both from
photon deflection and interferometry. Longer distances $L$ also offer some
possibility for improvement, but the dependence on $L$ is only as $L^{-1/2}$, so the
gain to be had in this area is not great. With greater distances $L$ also comes a
greater reliance on accurate cosmological models, which may legitimately be
questioned when we have introduced such an exotic modification of known physics as a
photon self coupling.

The considerations with the interferometric and the more important birefringence
bounds on $\mu$ are similar. A detailed study of the polarization angles of
quasars' synchrotron radiation relative to their jet directions would probably
improve the constraints on the photon magnetic moment. Such combined analyses have
already been performed as part of searches for Lorentz-violating vacuum
birefringence, but they are not directly applicable here. The existing analyses
looked for a birefringence proportional to the distance $L$ (times cosmological
corrections). To place a bound on $\mu$, the analysis would need to be redone,
looking for a rotation in the plane of the polarization proportional not to $L$, but
with its absolute value proportional to $\sqrt{L}$ (and having a large variance). The
$\sqrt{L}$ dependence is
characteristic of the magnitude of the integral of the magnetic field along the line
of sight, so this is the same dependence that would be expected from the Faraday
effect due to propagation through magnetized plasmas. Aggregated data sets have been
used to search for this effect and to place bounds on the extragalactic $\vec{B}$,
but since the Faraday effect is proportional to the wavelength squared, these
analyses are again not directly applicable.

Yet
while some improvement in the bounds given here is certainly possible, there is not a
great deal of motivation to push the best bound $q$ or $\mu$ down by one or two
orders of magnitude.
What should be interesting about the present work is that it offers a new way to
place bounds on these exotic possibilities. There are a number of well known reasons
why it is difficult to construct sensible theories of charged or magnetized photons.
This work can be seen as adding new
difficulties for such theories. We have introduced a new class of effects that arise
naturally in self-interacting electromagnetic theories, yet for which no evidence is
seen. This reinforces the idea that photon self couplings are not viable either
experimentally or theoretically.

In this paper, we have presented bounds on the photon charge and magnetic moment.
The charge constraints are at the $10^{-32}$ $e$ level or better if all photons
carry the same charge and the $10^{-46}$ $e$ level if oppositely charged photons
exist.
These bound come from the fact that Aharonov-Bohm phases do not interfere with the
interferometric imaging of distant galaxies. The best constraint on the photon
magnetic moment $\mu$ is at the $10^{-24}$ $e$ cm level and comes from the absence
of wavelength-independent photon birefringence. We know of no other previously
published bounds on this quantity. These results indicate that the common
assumption that photons have no self interactions is extremely well justified.

\section*{Acknowledgments}
The author is grateful to S. Crittenden for useful discussions and
V. V. Kobychev for information about recent work in this area.

\end{document}